\def \SAIT #1 #2 {{\em Mem.\ Soc.\ Astron.\ It.\/} {\bf #1}, #2}
\def \MESS #1 #2 {{\em The Messenger\/} {\bf #1}, #2}
\def \ASTRNACH #1 #2 {{\em Astron. Nach.\/} {\bf #1}, #2}
\def \AAP #1 #2 {{\em Astron. Astrophys.\/} {\bf #1}, #2}
\def \AAL #1 #2 {{\em Astron. Astrophys. Lett.\/} {\bf #1}, L#2}
\def \AAR #1 #2 {{\em Astron. Astrophys. Rev.\/} {\bf #1}, #2}
\def \AAS #1 #2 {{\em Astron. Astrophys. Suppl. Ser.\/} {\bf #1}, #2}
\def \AJ #1 #2 {{\em Astron. J.\/} {\bf #1}, #2}
\def \ANNREV #1 #2 {{\em Ann. Rev. Astron. Astrophys.\/} {\bf #1}, #2}
\def \APJ #1 #2 {{\em Astrophys. J.\/} {\bf #1}, #2}
\def \APJL #1 #2 {{\em Astrophys. J. Lett.\/} {\bf #1}, L#2}
\def \APJS #1 #2 {{\em Astrophys. J. Suppl.\/} {\bf #1}, #2}
\def \APSS #1 #2 {{\em Astrophys. Space Sci.\/} {\bf #1}, #2}
\def \ASR #1 #2 {{\em Adv. Space Res.\/} {\bf #1}, #2}
\def \BAIC #1 #2 {{\em Bull. Astron. Inst. Czechosl.\/} {\bf #1}, #2}
\def \JSQRT #1 #2 {{\em J. Quant. Spectrosc. Radiat. Transfer\/} {\bf #1}, #2}
\def \MN #1 #2 {{\em Mon. Not. R. Astr. Soc.\/} {\bf #1}, #2}
\def \MEM #1 #2 {{\em Mem. R. Astr. Soc.\/} {\bf #1}, #2}
\def \PLR #1 #2 {{\em Phys. Lett. Rev.\/} {\bf #1}, #2}
\def \PASJ #1 #2 {{\em Publ. Astron. Soc. Japan\/} {\bf #1}, #2}
\def \PASP #1 #2 {{\em Publ. Astr. Soc. Pacific\/} {\bf #1}, #2}
\def \NAT #1 #2 {{\em Nature\/} {\bf #1}, #2}
\def\etal{{\it et~al.}}

\documentstyle{memsait}
\input epsf.sty
\begin{opening}
\title{RAPID SPINDOWN OF SGR 1900+14 FOLLOWING ITS SUPERBURST}
\author{DAVID M. PALMER$^1$}
\institute{$^1$Los Alamos National Lab, Los Alamos, NM, USA}
\date{} 
\end{opening}

\begin{document}

\oddpagefooter{}{}{} 
\evenpagefooter{}{}{} 
\ 
\bigskip

\begin{abstract}
SGR 1900+14 had a brief episode of exceedingly rapid spindown immediately
following its 1998 Aug. 27 superburst.
On a timescale of hours, it increased its period by a part in $10^{4}$.
The corresponding $\dot{P} \sim 10^{-8}$ is orders of magnitude higher than
the typical quiescent rate of $\dot{P} \sim 6\times10^{-11}$.
\end{abstract}

\section{Introduction}

The spindown history of SGR 1900+14, shown in Figure 1, shows a discontinuity
between the last pre-superburst observations, in 1998 May-June, and measurements
made in the days following the superburst (Woods \etal, 1999a). 
The long-term average spindown rates
both before the superburst, and in the months following the superburst, are each
consistent with a typical spindown rate of $\dot{P} \sim 6\times10^{-11} s s^{-1}$.
However the extrapolations
of those two measurement periods are discrepant by $\sim5\times10^{-4} s$, or
$10^{-4}$ of the neutron star's $5.16 s$ period.

If the excess period change accumulated uniformly through the interval June--August,
this would require a spindown of roughly double the
typical value.  SGRs are known to have variable spindown rates and, indeed,
the last May-June observations are consistent with this higher rate.

Alternatively, if the excess period change occurred between the superburst
and the follow-up observations beginning the next day, 
this would require an enormous $\dot{P}$, orders of magnitude higher than
the typical rate.

Observations of the very strong
pulsations in the superburst's tail are of too short a duration to resolve
between the two periods.  However, by comparing the phase of those pulsations against
the extrapolated phase from the following weeks,
the rapid and sustained spindown scenarios can be distinguished.
This phase comparison indicates that the spindown occurred in a few hours following
the burst.

\section{Data and Analysis}

Pulsations from SGR1900+14 have been observed by ASCA (Hurley \etal, 1999),
BeppoSAX (Woods \etal, 1999b),
RossiXTE (Kouveliotou \etal 1998, Woods \etal, 1999a),
and the BSA radio observatory (Shitov, 1999).
These pulsations were found in the quiescent flux, exclusive of bursts.
The main peaks of the bursts themselves occur equally often,
and with equal intensities, at all phases (Palmer, 1999).

The tail of the August 27 superburst (Fig. 2) shows extremely strong,
complex pulsations in its tail.
These pulsations were seen by several instruments including
KONUS-WIND (Mazets \etal, 1999),
and PCA-RXTE, as leakage through its shield.  In addition,
the PCA also observed strong pulsations in the tail of a burst on August 29
during pointed observations.

RXTE observations from 1998 Aug. 30-Sept. 27 form a good baseline for
the post-burst spin ephemeris of the source.  Trial foldings to
find the maximal $\chi^2$ of the pulse profile yield a
`September Ephemeris'
of $t_0 =$ 1998 Sept. 1.0 TDB,  $P = 5.160197(1) s$,
$\dot{P} \sim 6.1(3)\times10^{-11} s s^{-1}$, where quoted uncertainties
are based on a maximum phase error of 0.1 cycles (2 phase bins).
Systematic effects due to changes in the pulse profile
do not allow confidence in reducing this phase uncertainty, even
though the formal error from counting statistics
is much lower.
The $\chi^2$ map
(Fig. 3) shows no significant sidelobes, precluding aliasing.

Figure 4a shows a synopsis of the 1998 Aug.-Sept. observations of the SGR.
Figure 4b shows the corresponding pulse profiles based on folding
using the September Ephemeris.  The quiescent pulse profiles (beginning
with the Aug. 28 observation the day after the superburst) all line up
fairly well.  However, the pulse profile from the tail of the superburst
(far left) shows a large phase shift.

The value of the phase shift is ambiguous, due to the change in pulse
profile between the burst tail and the quiescent pulsations during the
following days.  Figure 5 shows the pulse profile of superburst tail in more detail,
and compares it to the quiescent profile for the following two days.
The agreement of Konus and PCA light curve shape and phase during interval B
demonstrates that barycentering, done manually for Konus
and using the {\it fxbary} FTOOL (Blackburn, 1995)
for the PCA, was consistently applied.

Four main peaks in the tail pulsations are labelled--a more detailed
analysis shows even more substructure.  Identifying which peak
(if any) of the tail corresponds to the main peak on the following days
is ambiguous.  Peak 1, corresponding to the smallest phase
shift, has almost disappeared by the start of interval B, 47 seconds
after the start of the burst.  The other three peaks are all plausible,
but a phase shift of $\Delta \phi_3 = 0.42$ cycles, appropriate for Peak 3,
allows the shape of the quiescent pulse to neatly envelope the peaks
in Interval C.  This does not exclude
peaks 2 and 4, but using the phase shift for those peaks
(0.26 and 0.62 cycles, respectively) would lead to similar
conclusions, as is discussed below.

The sign of the phase shifts,
as quoted, is appropriate for rapid spindown (increasing period) between
the tail of the burst and data taken the next day.  The potential ambiguity
of an integral number of spins ({\it i.e.} $\Delta \phi_3 = 1.42$ cycles)
is eliminated below.

\section{Spindown Rates}

The spin ephemeris of SGR 1900+14 derived from the post-superburst
data does not accurately predict the phase of emission at the time of the superburst.
This phase error is a consequence of error in the predicted period,
accumulated over the interval between the superburst and the following
observations: $\Delta\phi = \int{\Delta P(t)} / P^2 dt$.
A constraint on the size of the period error inversely
constrains the duration over which it must have persisted, allowing us to
determine the timescale of the spindown.

The 1996 Sept. -- 1998 June long term average spindown rate is
$\sim 6\times 10^{-11} s s^{-1}$, as is the rate for 1998 Sept.
However, extrapolations in both directions to the time of the 1998 Aug. 27
superburst give a discrepancy of $\Delta P = 0.57$ ms.  If the forward
extrapolation accurately predicts the spin rate at the time
of the superburst, the backward extrapolated ephemeris will accumulate
phase error at an initial rate of one cycle per
$T_{\delta} = P^2/\Delta P = 13.1$ hours.

A phase error of $\Delta\phi_3 = 0.42$ could thus be accumulated in
a time $\tau_0 = T_{\delta}\Delta\phi = 5.4$ hours.  This corresponds to
the model where the neutron star continues to spin at the higher speed
for a time $\tau_0$ after the burst, then instantaneously slows down
bya part in $10^{-4}$.  As such, it provides a lower limit on the
time between the superburst and the completion of the spindown.

More reasonably, we can model the spindown as a constant $\dot{P}_{fast}$ over an interval
$\tau_1 = 2 T_{\delta} \Delta\phi = 10.8$ hours, which requires
$\dot{P}_{fast} = \Delta P / \tau_1 = 1.5\times 10^{-8} s s^{-1}$.  This spindown rate is
$250\times$ the typical quiescent $\dot{P}$.  Any other functional form for the spindown
requires a higher $\dot{P}_{fast}$ at least part of the time.

Duncan ({\it priv. comm.}) has a model in which the period shortly after the burst is
$P(t) = P_0 + \Delta{P}/(1 + \tau_D/t)$.  For this model, $\tau_D = 2.2$ hours would
fit the first day's phase shift, but there would be an additional 0.1 cycle shift
on the following day, which is not apparent in Fig. 5.

No functional form in which $\dot{P}$ decreases monotonically back to the
typical value after the burst is consistent with the extra-turn
interpretation with $\Delta\phi_{3alias} = 1.42$.

\section{Discussion}

If any phase from peaks 2 through 4 of the tail pulses corresponds to
the peak of the quiescent emission,
then SGR 1900+14
must have had a spindown rate orders of magnitude higher than its typical
$\dot{P}$ for a few hours after the superburst.

One possible objection to
this is that the tail emission may be beamed in a different body-centered
direction than the quiescent emission, causing a shift between spin phase and emission phase.
However, the large (but not super)
baby burst from SGR 1900+14 two days later, on Aug 29, also shows strong pulsations
in its tail.  These tail
pulses are in phase with the the quiescent pulses before and after this burst (Fig. 6).
It is therefore reasonable to assume that the superburst
tail emission and the quiescent emission are indeed beamed in the same direction.

A typical glitch in a radio pulsar is a spin rate increase event, thought to be produced
by sudden coupling of the neutron star crust to its superfluid interior which is still
rotating at an earlier, faster rate.  An Anomalous X-ray Pulsar, thought to be a similar
object to an SGR, has also shown such a spin-up glitch (Kaspi \etal, 2000).

The August 27 spindown event is different from a typical glitch in both sign
and magnitude ($\Delta P/P = 10^{-4}$ {\it vs.} $\sim -10^{-6}$ for a `giant glitch').
Since the magnetosphere is providing the braking force, it is moving more slowly
than the superfluid interior, and thus coupling cannot cause a spindown.
Slowing the spin by increasing the moment of inertia of the neutron star is
energetically infeasible ($\sim10^{49}$ erg).
Therefore, it is more likely that the SGR is shedding angular momentum.

If the SGR superburst generated a massive wind held in co-rotation by the magnetosphere
until it decouples at the light cylinder, then $\sim10^{20} g$ of this wind would
slow the neutron star by the required amount.  The energetic cost of lifting
this wind against gravity is $\sim10^{40}$ erg, or $\sim10^{-3}$ of the total flare
X-ray energy.  This level of torque on the neutron star can be supported by
magnetar-level magnetic fields of $10^{14} G$ with a distortion of less than $10^{-9}$ radian
at the surface of the star.  Line emission suggesting the presence of iron around the
SGR has been seen in the August 29 burst (Strohmayer \& Ibrahim, 2000),
which may be crust material blown
into circumstellar space as part of the wind.

\section{Conclusion}

For an interval of several hours following its superburst, SGR 1900+14 had a
spindown rate that was orders of magnitude higher than its typical rate
during quiescence.  This spindown is consistent with a wind
with reasonable mass and energy parameters.

\section{Acknowledgements}

Konus data is courtesy of the Konus team.  This research has made use of data obtained from the
High Energy Astrophysics Space Research Center (HEASARC), provided by NASA's
Goddard Space Flight Center.

\begin{figure}
\section{Figures}
\caption[h]{Period history of SGR 1900+14, 1998 May to 1999 January.
Adapted from Woods \etal, (1999a).}
\caption[h]{Tail emission following the Aug. 27 superburst.  Gaps in the PCA data during
interval A are instrumental, due to buffer overflows.}
\caption[h]{Pulsation detection {\it vs.} Period
and Period derivative of SGR 1900+14 during the interval
1998 Aug. 30--Sept. 27, based on RXTE-PCA observations.}
\caption[h]{SGR 1900+14 behavior, 1998 Aug. 27 - Sept. 27.  All
data is from the RXTE PCA except for the Konus superburst data on Aug 27.
{\em a)} Observations and count rates--vertical spikes above the typical
$\sim$10 counts/0.1s rate indicate baby bursts.
{\em b)} Pulse profile folded using the `September Ephemeris' for the
superburst tail (Konus, Aug. 27) and during quiescent times with
bursts excluded.}
\caption[h]{SGR 1900+14 superburst tail pulsations, for the intervals
shown in Fig. 2, compared to those of the following two days.  Data are from
{\em A} Konus, {\em B} Konus \& PCA shield leakage, {\em C} PCA shield leakage,
and {\em Aug 28{\rm\&}29} PCA pointed observations.  Time intervals chosen are based
on data availability: the PCA data was interrupted by buffer overflows
during interval A, and Konus data ends with interval B.}
\caption[h]{The 'Big Baby' burst of 1998 Aug. 29, as seen by the RXTE-PCA.
The pulsations in its tail are in phase with those before and after the burst.}
\end{figure}

\begin{figure}
\epsfxsize=15cm 
\epsfbox{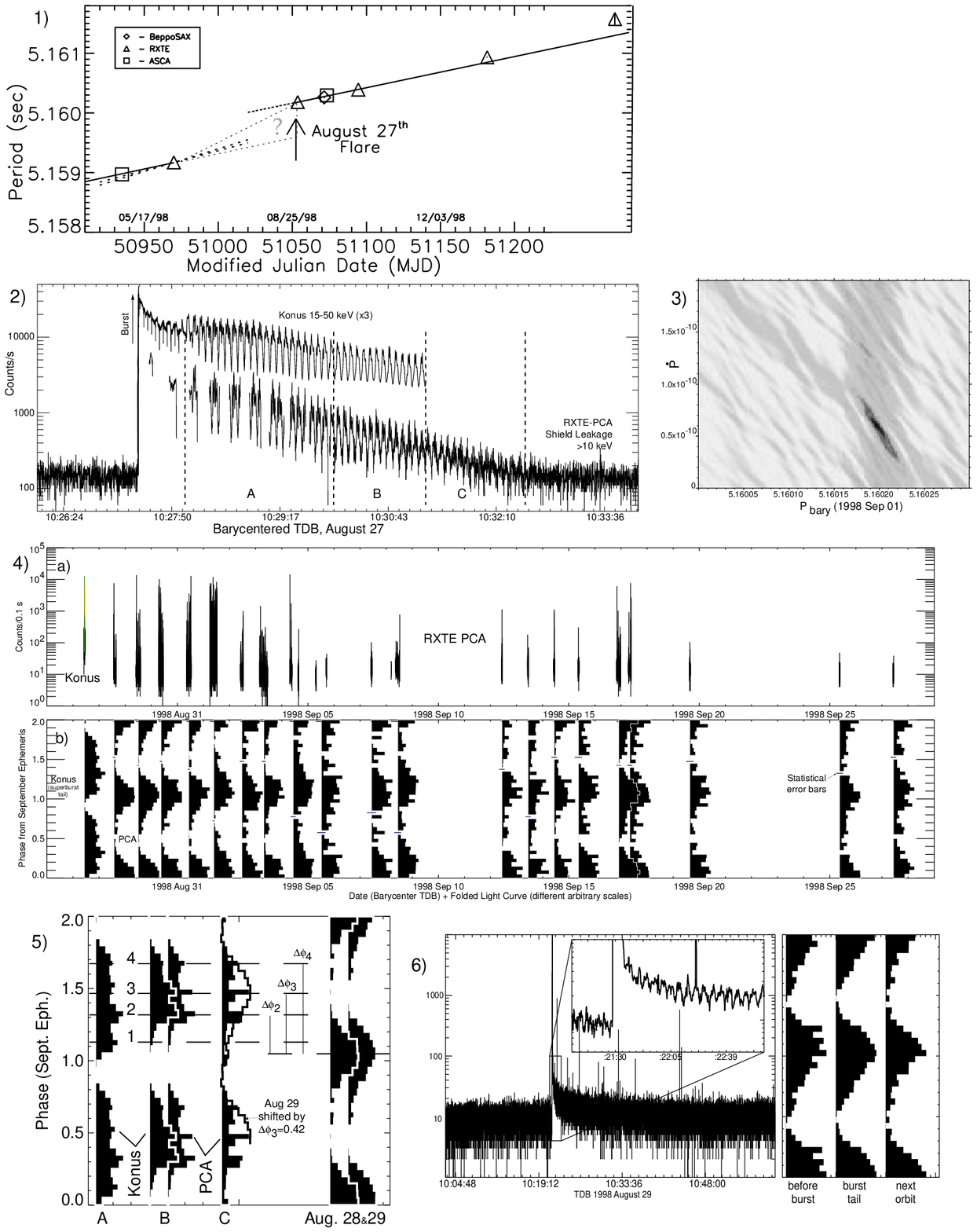} 
\end{figure}

\end{document}